\documentclass[prl,twocolumn,floats,aps,epsfig,nofootinbib,amssymb]{revtex4}
\usepackage{subfigure}
\usepackage{graphicx}
\usepackage{cancel}
\usepackage{amssymb}
\usepackage{textcomp}
\usepackage{amsmath}
\usepackage{bm}
\usepackage{times}
\usepackage{epsfig}
\usepackage{color}
\def\BR{\text{BR}}
\begin{document}
\title{\Large Testing the Mechanism for the LSP Stability at the LHC}
\author{Pavel Fileviez P{\'e}rez}
\author{Sogee Spinner}
\author{Maike K. Trenkel}
\address{
Phenomenology Institute, Department of Physics, University of Wisconsin-Madison, WI 53706, USA}
\date{\today}

\begin{abstract}
The lightest supersymmetric particle (LSP) is a natural candidate for the cold dark matter of the universe.  
In this letter we discuss how to test the mechanism responsible for the LSP stability at the LHC. 
We note that if $R$-parity is conserved dynamically one should expect a Higgs boson which decays mainly 
into two right-handed neutrinos (a ``leptonic" Higgs) or into two sfermions. The first case could exhibit 
spectacular lepton number violating signals with four secondary vertices due to the long-lived nature 
of right handed neutrinos. These signals, together with the standard channels for the discovery of SUSY, 
could help to establish the underlying theory at the TeV scale. 
\end{abstract}
\maketitle
{\bf Introduction:}
The Minimal Supersymmetric Standard Model (MSSM) is considered as one of the most appealing candidates 
for TeV scale physics since in this context one can protect the Higgs mass, there is a natural cold dark matter 
candidate of the universe if  $R$-parity is conserved, it can accommodate electroweak baryogenesis, 
and one can achieve the unification of gauge couplings at the high-scale. 
For a review of Supersymmetry (SUSY) see Ref.~\cite{SUSY-review}.

The signatures of low-energy SUSY at the LHC depends on the conservation or violation of $R$-parity. 
This symmetry is defined as $R=(-1)^{2S+3(B-L)}$, where S, B and L are the spin, baryon and lepton number, 
respectively. If $R$-parity is conserved the superpartners are produced in pairs and typical channels include 
multi jets, multi leptons and missing energy. The latter due to the LSP which might also be dark matter. 
When $R$-parity is broken single superpartner production is possible as well as observable lepton 
or baryon number violation at the LHC~\cite{RpV}.

The origin of $R$-parity conservation or violation can be understood in TeV scale theories based on the $B-L$ 
gauge symmetry. It is well-known that if the Higgs breaking $U(1)_{B-L}$ has even $B-L$ charge, 
$R$-parity is conserved after symmetry breaking~\cite{RpC}. Testing this mechanism at the LHC requires 
an investigation of the properties of the $Z_{BL}$ neutral gauge boson, the right-handed neutrinos 
needed to define an anomaly free theory, and the new $B-L$ Higgses. This is the main goal of this work. 

In this letter we study the properties of the Higgs sector of theories which explain the origin of $R$-parity 
conservation  and find that there are two major types of models. In the first type the Higgses allowing 
for the stability of  the LSP can decay mainly into two right-handed neutrinos (we call it 
the ``leptonic'' Higgs). We find that the $B-L$ Higgses could give rise to spectacular signals 
with four displaced vertices due to the long lifetime of right-handed neutrinos. 
See Fig. 1 for a naive representation of these signals. In the second type of model, the new 
Higgses can decay mainly into two sfermions where the final states depend on the SUSY spectrum. 
However, specific scenarios have peculiar signals with multi-leptons and multi-photons in the final states. 
Our results are relevant to understand the testability of the mechanism responsible for the LSP stability in 
low energy SUSY and together with the standard SUSY discovery channels, could 
help establish the underlying theory at the TeV scale.
%
\begin{flushleft}
\begin{figure}[tb]
\includegraphics[scale=1,width=5.0cm]{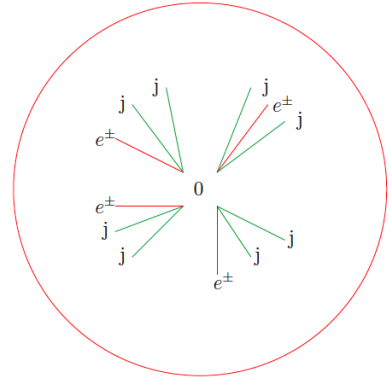}
\caption{Naive representation of the topology for the signals with secondary 
vertices and lepton number violation due to the existence of long lived right-handed Majorana neutrinos.}
\end{figure}
\end{flushleft}
%
{\bf Theoretical Framework:}
We begin by reviewing some details of the MSSM. As it is well-known, the 
superpotential is given by
\begin{equation}
{\cal W}_{MSSM}= {\cal W}_{RpC} \ + \ {\cal W}_{RpV},
\end{equation}
where ${\cal W}_{RpC}$ is the $R$-parity conserving part, 
\begin{equation}
{\cal W}_{RpC} = Y_u Q H_u u^c \ + \  Y_d Q H_d d^c \ + \  Y_e L H_d e^C \ + \ \mu H_u H_d,  
\end{equation}
and 
\begin{equation}
{\cal W}_{RpV} = \epsilon L H_u \ + \ \lambda L L e^c \ + \ \lambda^{'} Q L d^c \ + \  \lambda^{"} u^c d^c d^c,  
\end{equation}
contains the $R$-parity violating terms. The simplest way to imbed $R$-parity 
conservation into a gauge symmetry is through $B-L$ where the new gauge group is 
$$SU(3)_C \bigotimes SU(2)_L \bigotimes U(1)_Y \bigotimes U(1)_{B-L}$$
Now the terms in ${\cal W}_{RpV}$ are not allowed at the $B-L$ scale. Gauging $B-L$ requires the addition of three copies of right-handed neutrinos to satisfy anomaly cancellation, enhancing the superpotential to
\begin{equation}
{\cal W}_{B-L}= {\cal W}_{RpC} \ + \ Y_\nu L H_u \nu^c \ + \ {\cal W}_{extra}, 
\end{equation}
where the last term is dependent on the Higgs sector.  

There are two simple Higgs sector that allow for $R$-parity conservation:

\underline{Model I}: A pair of Higgses: $X,\bar{X} \sim (1,1,0,\pm 2)$ is added to the theory so that  the extra term in the above superpotential becomes
\begin{equation}
{\cal W}_{extra}^{(I)}=\mu_X X \bar{X} \ + \ f  \nu^C \nu^C X.
\end{equation}
In this model $B-L$ is broken by the vevs of $X$ and $\bar{X}$, and $R$-parity is 
an exact symmetry at the TeV scale. Due to the second term in the above equation, the neutrinos are Majorana fermions and the new physical Higgses, $X_1, X_2$ and $A_{BL}$, 
can decay at tree level into two right-handed neutrinos.  These decays are the key to revealing the properties of the $B-L$ breaking Higgses, \textit{i.e.} indicating if $R$-parity is really conserved at the TeV scale. In this case  
radiative symmetry breaking could explain the origin of symmetry breaking at the TeV scale~\cite{Fate}.

\underline{Model II}: $S,\bar{S} \sim (1,1,0, \pm n_{S})$ with even $n_{S} \neq 2$ , are introduced and the extra term in the superpotential is 
\begin{equation}
{\cal W}_{extra}^{(II)}=\mu_S S \bar{S}.
\end{equation}
Here the neutrinos are Dirac fermions and the new physical Higgses, $S_1, S_2$ and $A_S$, 
do not couple to SM fields at tree level. These Higgses can decay at tree level into two 
sfermions and give rise to peculiar signals in specific SUSY scenarios. It is important to mention 
that an odd $|n_{S}| \neq 1/2, 1$, will have higher-dimensional operators which will affect the stability of the LSP, although it may still live long enough to be a dark matter candidate.

{\bf Symmetry Breaking:} In order to simplify the symmetry breaking discussion
we introduce the generic fields, $\phi,\bar{\phi} \sim (1,1,0,\pm n_{\phi})$. 
Then, $\phi(\bar{\phi})$ can be $X(\bar{X})$  in Model I or $S(\bar{S})$ in Model II. 
The relevant soft terms for our discussion are:
\begin{eqnarray}
- {\cal L}_{Soft} &\subset & \left( a_\nu \tilde{L} H_u \tilde{\nu}^c \ - \ b_{\phi} \phi \bar{\phi} 
\ + \  \frac{1}{2} M_{BL} \tilde{B}^{'} \tilde{B}^{'}  \ + \  \text{h.c.} \right) 
\nonumber \\
& + &  m_\phi^2 |\phi|^2 \ + \ m_{\bar{\phi}} |\bar{\phi}|^2 \ + \ m_{\tilde{\nu}^c}^2 |\tilde{\nu}^c|^2,
\end{eqnarray}
where $\tilde{B}^{'}$ is the $B-L$ gaugino. Spontaneous $B-L$ breaking and $R$-parity conservation requires 
the vacuum expectation values (VEV) of $\phi$ and $\bar{\phi}$ to be non-zero. 
Using $\left<\phi \right>=v/\sqrt{2}$ and $\left<\bar{\phi}\right>=\bar{v}/\sqrt{2}$ for the VEVs, 
one finds
\begin{eqnarray}
V&=& \frac{1}{2}|\mu_{\phi}|^2 \left( v^2 + \bar{v}^2\right) \ - \ b_{\phi} v \bar{v} \ + \ \frac{1}{2}m_{\phi}^2 v^2 \ + \ \frac{1}{2} m_{\bar{\phi}}^2 \bar{v}^2 
\nonumber \\
 & + & \frac{g_{BL}^2}{32} n_{\phi}^2 \left( v^2 - \bar{v}^2 \right)^2.
\end{eqnarray}
Assuming that the potential is bounded from below
we get:
\begin{equation}
2 b_{\phi} < 2 |\mu_{\phi}|^2 + m_{\phi}^2 + m_{\bar{\phi}}^2,
\end{equation} 
while
\begin{equation}
b_{\phi}^2 > \left( |\mu_{\phi}|^2 + m_{\phi}^2 \right) \left( |\mu_{\phi}|^2 + m_{\bar{\phi}}^2 \right),
\end{equation}
is necessary for a nontrivial vacuum. Minimizing with respect to $v$ 
and $\bar{v}$ one gets
\begin{eqnarray}
|\mu_{\phi}|^2 + m_{\phi}^2 - \frac{1}{2} m_{Z_{BL}}^2 \cos 2 \beta^{'} - b_{\phi} \cot \beta^{'}=0,
\\
|\mu_{\phi}|^2 + m_{\bar{\phi}}^2 + \frac{1}{2} m_{Z_{BL}}^2 \cos 2 \beta^{'} - b_{\phi} \tan \beta^{'}=0,
\end{eqnarray}
with $\tan \beta^{'}=v/\bar{v}$ and $m_{Z_{BL}}^2=g_{BL}^2 n_{\phi}^2 (v^2 + \bar{v}^2)/4$.
Notice that the minimization conditions are quite similar to the MSSM conditions.

{\underline{Higgs Sector}}: The Higgs sector contains two extra CP-even neutral Higgs fields: $H_1$ and $H_2$, 
and one CP-odd Higgs, $A_{\phi}$, where this general notation will be replaced by the specific notation outlined in the model subsections.  The eigenvalues in the CP-even neutral Higgs sector read as
\begin{equation}
	m_{H_{1,2}}^2=\frac{1}{2} \left( m_{A_\phi}^2 + m_{Z_{BL}}^2 
	\mp \sqrt{D} \right),
\end{equation}
with
\begin{equation}
	D=(m_{A_\phi}^2 - m_{Z_{BL}}^2)^2  +  4 m_{Z_{BL}}^2 m_{A_\phi}^2 \sin^2 (2 \beta^{'}),
\end{equation}
and the mixing angle obeys the following relation:
\begin{equation}
	\frac{\tan 2 \alpha'}{\tan 2 \beta'} = \frac{m_{A_\phi}^2 +  m_{Z_{BL}}^2}{m_{A_\phi}^2 -  m_{Z_{BL}}^2}.
\end{equation}
Notice that in the limit, $m_{Z_{BL}}^2 \gg m_{A_\phi}^2$, one finds that
\begin{eqnarray}
m_{H_1}^2 &\sim& m_{A_\phi}^2 \left(  1 - \sin^2 2 \beta' \right), \\
m_{H_2}^2 &\sim & m_{Z_{BL}}^2 + m_{A_\phi}^2 \sin^2 2 \beta'. 
\end{eqnarray}
Then, assuming $g_{BL} \sim {\cal O}(1)$ and the experimental limit, $m_{Z_{BL}}/g_{BL} > 3 $ TeV~\cite{PDG}, 
one expects two light Higgses, $H_1$ and $A_{\phi}$, and a heavy one, $H_2$, when $m_{A_\phi}$ is small: a technically natural limit as a massless pseudoscalar corresponds to an enhanced global symmetry of the Lagrangian.

{\bf Production Mechanisms and Higgs Decays}:
Testing these scenarios for $R$-parity conservation requires
an understanding of how to produce the $B-L$ Higgses at the LHC and their decays.
The following production mechanisms are possible:
a) Single Production via Gluon fusion: $pp \to H_1$, b) Pair production: 
$pp \to Z_{BL}^* \to  H_1 A_{\phi}$, c) Associated production: 
$pp \to Z_{BL}^* \to Z_{BL} H_1$ and d) $Z_{BL}$ boson fusion.
The single production is quite model dependent while the associated production and 
the vector boson fusion are very small due to the experimental limits 
on the $Z_{BL}$ mass. Therefore, we focus mainly on the pair production, which also has the interesting property that while it is not a SUSY process, it is absent in minimal non-SUSY $B-L$ models.
We postpone the study of the other production channels for a future publication~\cite{Pavel-Sogee-Maike-2}. 

\begin{figure}[tb]
\includegraphics[scale=1,width=7.0cm]{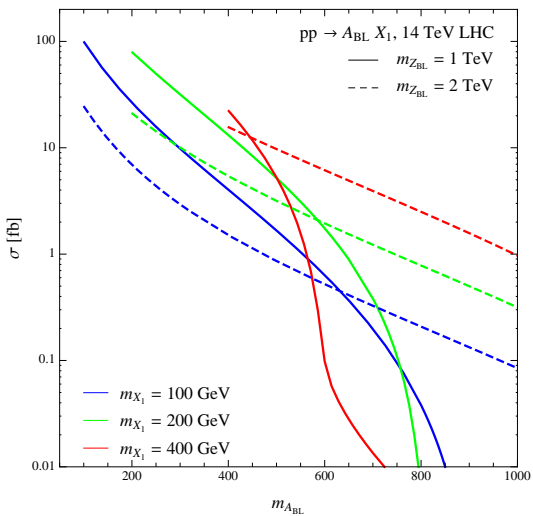}
\caption{$A_{BL}$ and $X_1$ pair production hadronic cross section in femtobarns 
for $14$ TeV center of mass energy at the LHC as a function of the mass $m_{A_{BL}}$.
We use the MSTW 2008 LO pdf at a central factorization scale $\mu = (m_{A_{BL}} + m_{X_1})/2$.}
\label{qqAX.cross.section}
\end{figure}

The pair production cross section for Model I, $pp \to X_1 A_{BL}$, 
versus the CP-odd scalar Higgs mass is shown in Fig.~2, using the MSTW2008 LO pdf at a central 
factorization scale $\mu = (m_{A_{BL}} + m_{X_1})/2$.
Here, for illustration, we show two different $Z_{BL}$ masses: 1 TeV (solid) and 2 TeV (dashed) (assuming $g_{BL}$ at its maximum allowed value in each case), 
and three different values for the Higgs mass, $m_{X_1}=100 \ \rm{GeV}, 200 \ \rm{GeV}$ 
and $400$ GeV.  This process also has a mild dependence on the SUSY spectrum through the $Z_{BL}$ width. All results are shown for center of mass energy of 14 TeV.  Larger cross sections for larger values of $m_{X_1}$ (for $m_{X_1} + m_{A_{BL}}$ below the $m_{Z_{BL}}$ threshold) correspond to larger value of $\tan \beta'$, a parameter that is uniquely determinable at each point on the curves.  For details see~\cite{Pavel-Sogee-Maike-2}.  This plot shows that when the pseudoscalar mass, $m_{A_{BL}}$, is smaller than 500 GeV, the hadronic cross 
section can be above 1 fb for these input parameters and in the most optimistic region ($m_{X_1} < 200$ GeV and $m_{A_{BL}} < 250$ GeV) can go as high as 100 fb indicating a large number of events.  Cross sections for Model II do not simply scale as $n_S^2$ as it also effects the width of $Z_{BL}$, see~\cite{Pavel-Sogee-Maike-2}.

With these values for the cross section we are now ready to study the possible 
decays in these models, focusing on tree-level two-body decays for $H_1$ and $A_{\phi}$ as the most accessible fields. 
In Model I the CP-even physical Higgs, $X_1$, can decay at tree level into two right-handed neutrinos, 
into two sfermions or into two $B-L$ neutralinos, while the CP-odd Higgs can decay into two right-handed neutrinos, 
two $B-L$ neutralinos, and $X_1$ and $Z_{BL}$. However, since the collider bounds on the $Z_{BL}$ mass 
are strong and we're focusing on a light Higgs,  the decays into 
two $Z_{BL}$s and sfermions are suppressed. Therefore, we are mainly interested in the decays 
into two right-handed neutrinos which would reveal the nature of the $B-L$ Higgses. 
Notice that in this case the right-handed neutrinos are Majorana fermions and their
decays produce lepton number violating signals. 

In Model II the physical Higgses, $S_1$, $S_2$ and $A_S$ do not have couplings to the 
fermions at tree level. Then, the CP-even Higgses can decay mainly into two sfermions or two $B-L$ neutralinos. The CP-odd Higgs can decay into $Z_{BL}$ and $S_1$ or two 
$B-L$ neutralinos. One possible interesting scenario corresponds to the case where $S_1$ can decay mainly into 
selectrons and $A_{S}$ through the three body decays, $A_{S} \to S_1 (Z_{BL}^*) \to S_1 e^+_i e^-_i$. The possible 
signals in this model depend of the particular SUSY spectrum. The simplest scenarios will be discussed in 
the next section.

\begin{figure}[tb]
\includegraphics[scale=1,width=7.0cm]{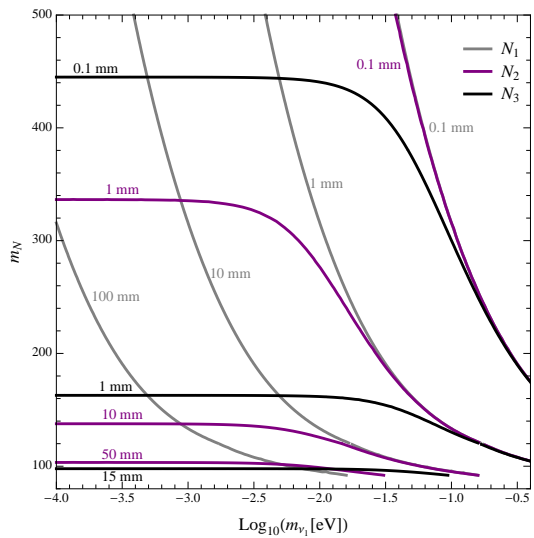}
\caption{Lines of constant decay length in millimeter for the right-handed neutrinos in the $m_N$--$m_{\nu_1}$ plane 
when one has a normal hierarchy spectrum for the left-handed neutrinos. We assume for simplicity the tri-bimaximal ansatz for neutrino mixings, 
$\Omega=\mathbf{1}$, and $m_h=120$ GeV.}
\end{figure}

Since in Model I the Higgses can decay mainly into two right-handed neutrinos, their properties are important when discussing signals at the LHC. The possible decays for N are: $N \to W^{\pm} e^{\mp}, Z \nu, \nu h_i$, where $h_i$ 
are the MSSM Higgses. The partial width for these decays are proportional to the mixing between the light SM 
neutrinos and N. The relevant mixing matrix is given by 
\begin{eqnarray}
V_{\ell N}= \ V_{PMNS} \ m^{1/2} \ \Omega \ M^{-1/2}_N.
\label{mixing1}
\end{eqnarray}
Here $V_{PMNS}$ is the PMNS mixing matrix, $m=diag(m_{\nu_1}, m_{\nu_2}, m_{\nu_3})$ are the physical neutrino masses, 
and $\Omega$ is a complex orthogonal matrix, which conveniently parameterizes some of the unknown degrees of freedom of 
the neutrino sector. In Fig. 3 we show lines of constant decay length for the right-handed neutrinos in the right-handed neutrino mass$-$lightest left-handed neutrino mass plane for a normal hierarchy (NH) spectrum for the light neutrinos.  Here we have used the central values for the atmosphere and solar mass squared difference, the tri-bimaximal ansatz and assumed $\Omega = \mathbf{1}$. As can be appreciated 
from Fig. 3, the right-handed neutrinos are long-lived in the full parameter space considered implying 
the existence of secondary vertices. For a detailed study of the right-handed decays see 
Ref.~\cite{TypeI-Zprime}.
 
{\bf Signals at the LHC}:
In Model I we will focus on the light Higgs bosons scenario, with $m_{X_1} > 2 m_{N}$, which decay only into two right-handed neutrinos. 
This corresponds to the most optimistic case where one has lepton number violation through the right-handed neutrino decays.  For single production of the $B-L$ Higgs boson $X_1$, the following lepton number violating signals at the LHC are possible:
\begin{align}
pp \to X_i \to N \ N  \to e^{\pm}_i W^{\mp} e^{\pm}_j W^{\mp} \to e^{\pm}_i  e^{\pm}_j  4j.
\nonumber 
\end{align}
Since the single production depends heavily on the SUSY spectrum we shift our focus to the pair production as the arena for testing the stability of the LSP. In this case one has
\begin{align}
	pp \to  Z_{BL}^* \to X_1 A_{BL} \to  N N N N \to  e^{\pm}_i  e^{\pm}_j e^{\pm}_k  e^{\pm}_l   8j,
	\nonumber
\end{align}
leading to four same-sign leptons and eight jets in the final state and allowing for observation of lepton number violation.
These signals are even more spectacular once we consider the fact that the right-handed neutrinos are long-lived, which lead to four secondary vertices.  The topology of these event are shown in Fig. 1.

We can perform a naive estimate to gain an appreciation for the number of events for this four same-sign leptons 
and 8 jets signal using a cross section of $100$ fb and an integrated luminosity of $100 \text{ fb}^{-1}$:
\begin{eqnarray}
N_{4 e 8 j}& \approx& \sigma (pp \to X_1 A_{BL}) \times \BR (X_1 \to N_1 N_1 ) 
\nonumber \\
&\times&  \BR (A_{BL} \to N_1 N_1 ) \times 2 \BR(N_1 \to e^+ W^- )^4  
\nonumber \\
&\times & \BR (W \to jj)^4  \times {\cal L} 
\nonumber \\
 &\approx& 100 fb \times (1/3) \times (1/3) \times 2 (3/10)^4  \times (6/9)^4 
 \nonumber \\
 && \times 100 fb^{-1} \approx 4.
\end{eqnarray}
Where the $\text{BR}\left(X_1 \to N_1 N_1\right) = 1/3$ due to the three possible generations of right-handed neutrinos and $\text{BR}\left(N_1 \to e^+ W^-\right) \approx 3/10$ due to model specific parameters and kinematics, see~\cite{Pavel-Sogee-Maike-2} and the benchmark defined below.  In order to make a realistic calculation of the number of events we pick a benchmark scenario:

Benchmark Scenario I: $m_{A_{BL}} = 220$ GeV, $m_{X_1} = 200$ GeV, $m_{Z_{BL}} = 1$ TeV, 
$\mu_{BL} = 150$ GeV, $M_{BL} = 150$ GeV, $m_{N_i} = 95$ GeV, for $i=1..3$, 
$m_{{\tilde{t}}_1}  =150$ GeV, $m_{\tilde{\tau}_1} = 150$ GeV and all other sfermions at 1 TeV.

The sfermion masses effect the $Z_{BL}$ width.  In this case  
$\sigma_{pp \to X_1  A_{BL}} =65.7$ fb,
and we display the predicted number of events in Table I for the five possible final states with an e and/or a $\mu$.  
We also show the combinatorics factor which takes into account the branching ratios of the Higgses 
into right-handed neutrinos, right-handed neutrinos into leptons and $W$s into jets.  
This number can be multiplied by any cross section and integrated luminosity to yield 
the number of events. 
\begin{table}[htdp]
\begin{center}
\begin{tabular}{|c||c|c|}
\hline
Final State	& Combinatorics     &  $\ $ Number of Events $\ $
 \\
\hline
\hline
$4 \, e^\pm \, 8 \, j$	&	0.00072	& 4.8
\\
$ 3\, e^\pm \, \mu^\pm\, 8 \, j$ & 0.0012 & 7.6
\\
$ \ 2\, e^\pm \, 2 \mu^\pm\, 8 \, j \ $ & 0.0015 & 9.7
\\
$ \, e^\pm \, 3 \mu^\pm\, 8 \, j$ & 0.00081 & 5.3
\\
$4 \mu^\pm\, 8 \, j$ & 0.00035 & 2.3
\\
\hline
\end{tabular}
\caption{\small{Number of events for the five possible four same-sign leptonic final states (with $e$ or $\mu$) for a luminosity 
of $100 \ \text{fb}^{-1}$ and a pair production cross section of $65.7$ fb corresponding to benchmark I.  We also display 
the combinatorics factor which combines the branching ratios for the Higgses into right-handed neutrinos, right-handed neutrinos 
to specific leptonic final states and $W$s into jets.}}
\end{center}
\label{NE}
\end{table}
In this case, if we ignore the displaced vertices, the main SM background is $t \bar{t} W^{\pm} t \bar{t} W^{\pm}$ and 
it has a negligible cross section so that while there are only a few number of events, they are background free. This does not change the fact though that 
the reconstruction would be quite challenging due to the presence of eight jets in the final state. Imposing the condition 
that the invariant mass of two jets, $|M(jj)-M_W | < 15 $ GeV \cite{TypeI-Zprime}, can improve the reconstruction process 
as well as the order millimeter displaced vertices due the long lifetimes of the right-handed neutrinos.

In Model II the Higgs, $S_1$, can decay mainly into two sfermions. Here, for simplicity we focus on a scenario 
where the gravitino is the LSP with a simplified spectrum: $m_{\tilde{G}} < m_{\tilde{\chi}_1} \ < \ m_{\tilde{e}_i},  m_{\tilde{q}_i}, m_{\tilde{\nu}_i} \ < \ m_{S_1}/2 $. This type of spectrum could be obtained in gauge mediation. Assuming
that the neutralino is Bino-like, the pair production could lead to signals with mutileptons, multiphotons and 
missing energy: 
\begin{align}
pp \to S_1 A_S \to \tilde{e}^* \tilde{e} \  S_1 e^+_i e^-_i  \to  e^{\pm} e^{\mp} e^{\pm} e^{\mp} e^+_i e^-_i \gamma \gamma \gamma \gamma + E_T^{miss}
\nonumber
\end{align}
In principle, these type of signals are quite spectacular because they could also include displaced vertices due to the lifetime of the neutralinos. Unfortunately, the branching ratio of the 
Higgs into sfermions is quite spectrum dependent. For a recent study of  long-lived neutralinos see Ref.~\cite{Meade}. 
We postpone the study of these channels for a future publication.

{\bf{Summary and Outlook}}:
In this letter we have discussed how to test the mechanism responsible for the LSP stability at the LHC. 
We note that if $R$-parity is conserved dynamically, a Higgs boson which decays mainly 
into two right-handed neutrinos (``leptonic" Higgs) or into two sfermions is likely. In the first case one could have 
spectacular lepton number violating signals with four secondary vertices due to the existence 
of long-lived right-handed neutrinos. The second case could have peculiar signals with multileptons and multiphotons.
A more detailed study will appear in a future publication~\cite{Pavel-Sogee-Maike-2}.  
These signals, together with the standard channels for the discovery of SUSY, could help us to establish the 
underlying theory at the TeV scale. 

{\textit{Acknowledgments}}:
{\small P.F.P. would like to thank Espresso Royale for hospitality. 
We would like to thank V. Barger and T. Han for discussions. This work is supported in part by the U.S. Department of Energy under 
grant No. DE-FG02-95ER40896, and by the Wisconsin Alumni Research Foundation.}

\end{document}